\documentclass[english,superscriptaddress,showpacs,manuscript]{revtex4-1}
\pdfoutput=1
\usepackage[T1]{fontenc}
\usepackage[latin9]{inputenc}
\usepackage{float}
\usepackage{mathtools}
\usepackage{bm}
\usepackage{amsmath}
\usepackage{amssymb}
\usepackage{graphicx}
\usepackage{esint}

\makeatletter

\newcommand{\lyxmathsym}[1]{\ifmmode\begingroup\def\b@ld{bold}
  \text{\ifx\math@version\b@ld\bfseries\fi#1}\endgroup\else#1\fi}

\@ifundefined{textcolor}{}
{%
 \definecolor{BLACK}{gray}{0}
 \definecolor{WHITE}{gray}{1}
 \definecolor{RED}{rgb}{1,0,0}
 \definecolor{GREEN}{rgb}{0,1,0}
 \definecolor{BLUE}{rgb}{0,0,1}
 \definecolor{CYAN}{cmyk}{1,0,0,0}
 \definecolor{MAGENTA}{cmyk}{0,1,0,0}
 \definecolor{YELLOW}{cmyk}{0,0,1,0}
}

\makeatother

\usepackage{babel}
\begin{document}
\global\long\def\vc#1{\bm{#1}}

\title{Ab-initio electron scattering cross-sections and transport in liquid
xenon}

\author{G. J. Boyle}

\affiliation{College of Science, Technology \& Engineering, James Cook University,
Townsville 4810, Australia}

\author{R. P. McEachran}

\affiliation{Plasma and Positron Research Laboratory, Research School of Physical
Sciences and Engineering, Australian National University, Canberra,
ACT 0200, Australia}

\author{D. G. Cocks}

\affiliation{College of Science, Technology \& Engineering, James Cook University,
Townsville 4810, Australia}

\author{M. J. Brunger}

\affiliation{School of Chemical and Physical Sciences, Flinders University, Adelaide,
SA 5001, Australia }

\affiliation{Institute of Mathematical Sciences, University of Malaya, 50603 Kuala
Lumpur, Malaysia}

\author{S. J. Buckman}

\affiliation{Plasma and Positron Research Laboratory, Research School of Physical
Sciences and Engineering, Australian National University, Canberra,
ACT 0200, Australia}

\affiliation{Institute of Mathematical Sciences, University of Malaya, 50603 Kuala
Lumpur, Malaysia}

\author{S. Dujko}

\affiliation{Institute of Physics, University of Belgrade, Pregrevica 118, 11080
Belgrade, Serbia}

\author{R. D. White}

\affiliation{College of Science, Technology \& Engineering, James Cook University,
Townsville 4810, Australia}
\begin{abstract}
Ab-initio fully differential cross-sections for electrons scattering
in liquid xenon are developed from a solution of the Dirac-Fock scattering
equations, using a recently developed framework \cite{Boyletal15}
which considers multipole polarizabilities, a non-local treatment
of exchange, and screening and coherent scattering effects. A multi-term
solution of Boltzmann's equation accounting for the full anisotropic
nature of the differential cross-section is used to calculate transport
properties of excess electrons in liquid xenon. The results were found
to agree to within $25$\% of the measured mobilities and characteristic
energies over the reduced field range of $10^{-4}$--$1$ Td. The
accuracies are comparable to those achieved in the gas phase. A simple
model, informed by highly accurate gas-phase cross-sections, is presented
to  improve the liquid cross-sections, which was found to enhance
the accuracy of the transport coefficient calculations.
\end{abstract}
\maketitle

\section{Introduction}

An understanding of the behavior of free electrons in liquids and
dense systems is of interest to both fundamental physics research
and to technological applications \cite{Boyletal15}. In particular,
liquid-phase noble gases are used in high-energy particle detectors
and at present several LAr (liquid argon) and LXe (liquid xenon) TPCs
(time projection chambers) have been built for dark matter searches
\cite{Regenfus2010,Aprile2010,Aprile2011g,Alner2007g,baller2014liquid},
neutrino detection \cite{ICARUS2004,ICARUS2011,baller2014liquid,Aprile2010},
and have also been used in high-energy beam-line experiments \cite{ATLAS2014,baller2014liquid}.
Optimizing the performance of these liquid TPC particle detectors
requires an accurate understanding of electron drift and diffusion
in noble liquids subject to electric fields. In a previous paper,
Boyle et al. \cite{Boyletal15}, we investigated the elastic scattering
of electrons from gas-phase and liquid-phase argon. In this paper,
we extend the previous discussion to consider elastic scattering of
electrons in gas-phase and liquid-phase xenon, using the same techniques
previously outlined \cite{Boyletal15}.

The study of excess electrons in dense gases and liquids involves
many effects that are not significant in dilute gaseous systems. When
the de Broglie wavelength of the electrons (near thermal energies)
is comparable to the interatomic spacing of the medium, scattering
occurs off multiple scattering centres simultaneously, rather than
through binary scattering. Furthermore, these scattering centres are
highly correlated in space and time. Historically, transport simulations
neglected these correlations and simply scaled calculations in the
dilute gas phase to higher densities. It has been shown \cite{Boyletal15}
that this simplistic approach cannot explain the non-linearities seen
in the experiments . As we have described in Boyle et al. \cite{Boyletal15},
there are other theoretical approaches to exploring the effect of
liquid correlations on the transport of light particles \cite{GeeFree86,Braglia82,Borghesani2006,Sakai2007,White2009,Kunhardt1991},
however these either require empirical inputs, are applicable only
close to equilibrium, or have heuristically combined the liquid effects
identified above to obtain an effective cross-section.

As also discussed in Boyle et al. \cite{Boyletal15}, we employ the
\textit{ab initio} procedure of Cohen and Lekner~\cite{Cohen1967}
using updated scattering theory to address transport in dense systems
under a kinetic theory framework. Atrazhev et al. \cite{Atrazhev1981}
have used a similar procedure based on a simplification of the Cohen
and Lekner theory where they argued that, for small energies, the
effective cross-section becomes dependent on the density only and
they obtained good agreement with experiment. However, the distance
at which to enforce this new behaviour of the effective cross-section
remains a free parameter in the theory and this constant effective
cross-section must be found empirically. By performing a detailed
analysis of the partial phase shifts, Atrazhev and co-workers \cite{Atrazhev1996}
were able to isolate the important properties of the potential which
are required for accurate determination of the transport properties.
Our calculations instead avoid these difficulties by using accurate
forms for the electron-atom interaction. The transport theory employed
in this manuscript also represents an improvement over previous calculations
as we use a full multi-term treatment of the velocity distribution
function \cite{White2009,White2011} which utilizes all the available
anisotropic details of the scattering cross-sections. It is well known
that the often-used two-term approximation for transport calculations
can be in serious error \cite{White2003a}, and we compare our solutions
for transport in the xenon system using two-term and multi-term treatments.

In the following sections we consider the calculation of the macroscopic
swarm transport properties, in the gaseous and liquid xenon environments,
from the microscopic cross-sections, modified by the screening and
coherent scattering effects discussed above. We first discuss a multi-term
solution of Boltzmann's equation in Section \ref{TransportTheory-1},
using the calculation of the elastic scattering cross-section for
electrons in dilute gaseous xenon via the Dirac-Fock scattering equations
in Section \ref{ScatteringGas}. We determine the xenon pair-correlator
at the xenon triple point in Section \ref{PairCorrelation} using
a Monte-Carlo simulation of a Lennard-Jones with parameters fitted
from experimental data in the gas phase and compare the results with
previous calculations. The pair-correlator allows us to determine
an effective liquid scattering cross-section for xenon (see Section
\ref{sub:Liquid-xenon-cross-section}), in the manner outlined in
Boyle et al. \cite{Boyletal15}. The application of these cross-sections
to determine macroscopic transport properties via kinetic theory is
also outlined in Section \ref{TransportTheory-1}. We present the
results of our transport calculations in Section \ref{sec:Results}.
Initially in Section \ref{sub:Electrons-in-gaseous} we consider only
electrons in gas-phase xenon, and we compare the reduced mobility
and characteristic energies with swarm experiment measurements for
a range of fields. Electrons in liquid-phase xenon are then considered
in Section \ref{sub:Electrons-in-liquid}. The impact of anisotropic
scattering and validity of the two-term approximation is also investigated,
and we highlight the importance of using a multi-term framework when
solving Boltzmann's equation accurately. Finally, we discuss a scaling
procedure to adapt other gas-phase cross-section sets, such as those
compiled from various theoretical and experimental sources, to the
liquid phase in Section \ref{sub:Electrons-in-liquid-2}, and compare
the transport properties. Throughout this paper we will make use of
atomic units ($m_{e}=e=a_{0}=\hbar=1$) unless otherwise specified.

\section{Multi-term solution of Boltzmann's equation\label{TransportTheory-1}}

A dilute swarm of electrons moving through gaseous or liquid xenon,
subject to an external electric field $E$, can be described by the
solution of the Boltzmann's equation for the phase-space distribution
function $f(\vc r,\vc v,t)$~\cite{Boltzmann1872}: 
\begin{equation}
\frac{\partial f}{\partial t}+\vc v\cdot\nabla f+\frac{e\vc E}{m_{e}}\cdot\frac{\partial f}{\partial\vc v}=-J(f),\label{eq:BE}
\end{equation}
where $\vc r,$ $\vc v$ and $e$ denote the position, velocity and
charge of the electron respectively. The collision operator $J(f)$
accounts for all the necessary collision types and interactions between
the electrons of mass $m_{e}$ and the background medium. In this
paper we will only consider reduced electric field strengths $E/N$
(where $N$ is the number density of the background material) such
that there is no significant contribution from excitation collisions
in xenon. 

In swarm experiments only a few macroscopic variables can be controlled
and/or measured~\cite{HuxlCrom74}. The most commonly reported transport
quantities for electrons in liquids are the mobility $\mu$ ($=W/E$,
where $W$ is drift velocity), and the transverse and longitudinal
characteristic energies, $D_{T}/\mu$ and $D_{L}/\mu$ respectively.
As shown in \cite{Boyletal15}, we can calculate these coefficients
through multi-term spherical harmonic representations of the necessary
velocity distribution functions: 

\begin{equation}
F(\vc v)=\sum_{l=0}^{\infty}\, F_{l}(v)P_{l}^{0}(\cos\theta)\;,\label{sphericalharmonic-1-1}
\end{equation}

\begin{equation}
F^{(L)}(\vc v)=\sum_{l=0}^{\infty}\, F_{l}^{(L)}(v)P_{l}(\cos\theta)\;,\label{sphericalharmonic-1-1-1}
\end{equation}
and 

\begin{equation}
F^{(T)}(\vc v)=\sum_{l=0}^{\infty}\, F_{l}^{(T)}(v)P_{l}^{1}(\cos\theta)\;.\label{sphericalharmonic-1-1-1-1}
\end{equation}
Note that $\theta$ denotes the angle relative to the electric field
direction (taken to be the $z$-axis) and $P_{l}^{m}(\cos\theta)$
are associated Legendre polynomials. The Boltzmann equation can be
re-written as the following hierarchy of equations~for these expansion
coefficients:
\begin{equation}
J^{l}F_{l}+\frac{l+1}{2l+3}a\left(\frac{\partial}{\partial v}+\frac{l+2}{v}\right)F_{l+1}+\frac{l}{2l-1}a\left(\frac{\partial}{\partial v}-\frac{l-1}{v}\right)F_{l-1}=0,\label{eq:homogenous}
\end{equation}

\begin{equation}
J^{l}F_{l}^{(L)}+\frac{l+1}{2l+3}a\left(\frac{\partial}{\partial v}+\frac{l+2}{v}\right)F_{l+1}^{(L)}+\frac{l}{2l-1}a\left(\frac{\partial}{\partial v}-\frac{l-1}{v}\right)F_{l-1}^{(L)}=v\left(\frac{l+1}{2l+3}F_{l+1}+\frac{l}{2l-1}F_{l-1}\right),\label{eq:longitudinal}
\end{equation}
and

\begin{equation}
J^{l}F_{l}^{(T)}+\frac{l+2}{2l+3}a\left(\frac{\partial}{\partial v}+\frac{l+2}{v}\right)F_{l+1}^{(T)}+\frac{l-1}{2l-1}a\left(\frac{\partial}{\partial v}-\frac{l-1}{v}\right)F_{l-1}^{(T)}=v\left(\frac{F_{l-1}}{2l-1}-\frac{F_{l+1}}{2l+3}\right),\label{eq:Transverse}
\end{equation}
where the $J^{l}$ represent the Legendre projections of the collision
operator detailed below and $a=eE/m_{e}$. We enforce the normalisation
condition: 
\begin{equation}
4\pi\int_{0}^{\infty}F_{0}(v)v^{2}dv=1\;.
\end{equation}
The solution of this hierarchy of equations then yields the mobility:
\begin{equation}
\mu=\frac{1}{E}\frac{4\pi}{3}\intop_{0}^{\infty}v^{3}F_{1}dv,
\end{equation}
and the characteristic energy: 

\begin{equation}
D_{\left(L,T\right)}=\frac{4\pi}{3}\intop_{0}^{\infty}v^{3}F_{1}^{(L,T)}dv\;.
\end{equation}
Note, this theory avoids the traditional two-term approximation used
in electron transport in liquids \cite{Cohen1967,Borghesani2006,Sakai2007a},
and is a true multi-term solution of Boltzmann's equation, whereby
the upper bound in each of the $l$-summations is truncated at a value
$l_{\mathrm{max}}$, and this value is incremented until some convergence
criteria is met on the distribution function or its velocity moments. 

We only consider low-energy elastic scattering in this study. Hence,
the collision operator appearing in (\ref{eq:BE}), which describes
the rate of change of the distribution function due to interactions
with the background material, will include elastic collisions only.
For the liquid systems considered here, the de Broglie wavelength
of the electron is often of the order of the average inter-particle
spacing $\sim N^{-1/3}$. In this energy regime, the electron is best
viewed as a wave that simultaneously interacts with multiple scattering
centres that comprise the medium. For liquid xenon, the average interparticle
spacing is approximately $2.6\,\lyxmathsym{\AA}$, implying that ``low''
energies are those less than $\sim0.5$~eV, which is several orders
of magnitude larger than the thermal energy of $\sim0.014$~eV. The
interaction potential is then modified from the dilute gas phase as
discussed in Section \ref{sub:Liquid-xenon-cross-section}. 

The Legendre projections of the elastic collision operator, in the
small mass ratio limit, accounting for coherent scattering, are given
by:

\begin{equation}
J^{0}\left({\Phi_{l}}\right)=\frac{m_{e}}{Mv^{2}}\frac{d}{dv}\left\{ v\nu_{1}(v)\left[v{\Phi_{l}}+\frac{kT}{m_{e}}\frac{d}{dv}{\Phi_{l}}\right]\right\} \label{eq:Davydov}
\end{equation}
\begin{equation}
J^{l}{\Phi_{l}}=\tilde{\nu}_{l}(v){\Phi_{l}}\qquad\text{for \ensuremath{l}\ensuremath{\ge}1},\label{eq:structurecollisionfreq}
\end{equation}
where $M$ is the mass of a xenon atom, $\Phi_{l}=\left\{ F_{l},F^{(L)},F^{(T)}\right\} $
and

\begin{equation}
\nu_{l}(v)=Nv2\pi\int_{0}^{\pi}\sigma(v,\chi)\left[1-P_{l}(\cos\chi)\right]\sin\chi d\chi,\label{eq:collisionfreq_gas}
\end{equation}
is the binary transfer collision frequency in the absence of coherent
scattering effects with $\sigma(v,\chi)$ the differential scattering
cross-section. In addition,

\begin{equation}
\tilde{\nu}_{l}(v)=Nv\left(2\pi\int_{0}^{\pi}\Sigma(v,\chi)\left[1-P_{l}(\cos\chi)\right]\sin\chi d\chi\right).\label{eq:collisionfreqs-1}
\end{equation}
are the structure-modified higher-order collision frequencies that
account for coherent scattering through:

\begin{equation}
\Sigma(v,\chi)=\sigma(v,\chi)\; S\left(\frac{2m_{e}v}{\hbar}\sin\frac{\chi}{2}\right),
\end{equation}
which represents an effective differential cross-section. $S$ is
the static structure factor, which can be determined from the pair-correlator
as discussed in Section \ref{sub:Liquid-xenon-cross-section}. In
what follows we also define the momentum transfer cross-sections without
and with coherent scattering via $\nu_{1}(v)=Nv\sigma_{m}(v)$ and
$\tilde{\nu}_{1}(v)=Nv\Sigma_{m}(v)$, respectively.

\section{Scattering of electrons by xenon gas\label{ScatteringGas}}

The theoretical procedures used in this paper, to describe the elastic
scattering of electrons from xenon atoms at low energies, are essentially
the same as those used in Boyle et al. \cite{Boyletal15} for electron
scattering from argon. We thus present only a short summary here and
refer the reader to reference \cite{Boyletal15} for more details. 

In the pure elastic scattering energy region, only two interactions
need to be considered, namely polarization and exchange. The polarization
interaction is accounted for by means of a long-range multipole polarization
potential, while the exchange interaction is represented by a short-range
non-local potential formed by antisymmetrizing the total scattering
wavefunction. The scattering of the incident electrons, with wavenumber
$k$, by xenon atoms can then be described in the gaseous phase by
the integral equation formulation of the partial wave Dirac-Fock scattering
equations (see Chen et al. \cite{Chen08} for details).

In matrix form, these equations can be written as: 

\begin{equation}
\left(\begin{array}{c}
f_{\kappa}(r)\\
g_{\kappa}(r)
\end{array}\right)=\begin{pmatrix}\begin{array}{c}
v_{1}(kr)\\
v_{2}(kr)
\end{array}\end{pmatrix}+\frac{1}{k}\int_{0}^{r}\mathrm{d}x\, G(r,x)\,\biggl[U(x)\begin{pmatrix}\begin{array}{c}
f_{\kappa}(x)\\
g_{\kappa}(x)
\end{array}\end{pmatrix}-\begin{pmatrix}\begin{array}{c}
W_{Q}(\kappa;x)\\
W_{P}(\kappa;x)
\end{array}\end{pmatrix}\biggr],\label{eq:DiracFock}
\end{equation}
where the local potential $U(r)$ is given by the sum of the static
and local polarization potentials i.e., 
\begin{equation}
U(r)=U_{\mathrm{s}}(r)+U_{\mathrm{p}}(r)\label{eq:potential}
\end{equation}
 and $W_{P}(\kappa;r)$ and $W_{Q}(\kappa;r)$ represent the large
and small components of the exchange interaction.

The precise form of these exchange terms is given in equation~(5)
of Boyle et al. \cite{Boyletal15}. In particular, the polarization
potential $U_{\mathrm{p}}(r)$ in equation~(\ref{eq:potential})
was determined using the polarized orbital method \cite{mceachran77}
and contained several static multipole terms as well as the corresponding
dynamic polarization term \cite{McEachran90,Mimnagh93}. In total,
the potential $U(r)$ contained all terms up to and including those
that behave as $r^{-16}$ asymptotically.

In equation (\ref{eq:DiracFock}), $f_{\kappa}(r)$ and $g_{\kappa}(r)$
are the large and small components of the scattering wavefunction.
Here the quantum number $\kappa$ is related to the total and orbital
angular momentum quantum numbers $j$ and $l$ according to $\kappa=-l-1$
when $j=l+1/2$ (spin-up) and $\kappa=l$ when $j=l-1/2$ (spin down).
Furthermore, $G(r,x)$ is the free particle Green's function given
in equations (23) and (24a,b) of Chen et al. \cite{Chen08}. The kinetic
energy $\epsilon$ of the incident electron and its wavenumber $k$
are related by: 
\begin{equation}
k^{2}=\frac{1}{\hbar^{2}c^{2}}\,\epsilon\,\bigl(\epsilon+2mc^{2}\bigr),\label{eq:kineticenergy}
\end{equation}
 where $c$ is the velocity of light which, in atomic units, is given
by $c=1/\alpha$ where $\alpha$ is the fine-structure constant.

In the integral equation formulation, the scattering phase shifts
can be determined from the asymptotic form of the large component
of the scattering wavefunction i.e., 
\begin{equation}
f_{\kappa}(r)_{r\rightarrow\infty}\kern-15pt \longrightarrow A_{\kappa}\,\hat{j}_{l}(kr)-B_{\kappa}\,\hat{n}_{l}(kr),\label{eq:wavefunction}
\end{equation}
 where $\hat{j}_{l}(kr)$ and $\hat{n}_{l}(kr)$ are the Riccati-Bessel
and Riccati-Neumann functions while the constants $A_{\kappa}$ and
$B_{\kappa}$ are defined in equations (7) and (8) of Boyle et al.
\cite{Boyletal15} . The partial wave phase shifts are then given
by:
\begin{equation}
\tan\delta^{\pm}(k)=\frac{B_{\kappa}}{A_{\kappa}},\label{eq:partials}
\end{equation}
 where the $\delta^{\pm}$ are the spin-up $(+)$ and spin-down $(-)$
phase shifts. The total elastic and momentum transfer cross-sections
are given, in terms of these phase shifts, by: 
\begin{equation}
\sigma_{{\rm el}}(k^{2})=\frac{4\pi}{k^{2}}\sum_{l=0}^{\infty}\Bigl\{(l+1)\,\sin^{2}\delta_{l}^{+}(k)+l\,\sin^{2}\delta_{l}^{-}(k)\Bigr\}\label{eq:elastic}
\end{equation}
 and 
\begin{multline}
\sigma_{{\rm mt}}(k^{2})=\frac{4\pi}{k^{2}}\sum_{l=0}^{\infty}\Bigl\{\frac{(l+1)(l+2)}{2l+3}\,\sin^{2}\bigl(\delta_{l}^{+}(k)-\delta_{l+1}^{+}(k)\bigr)+\frac{l(l+1)}{2l+1}\,\sin^{2}\bigl(\delta_{l}^{-}(k)-\delta_{l+1}^{-}(k)\bigr)\\
+\frac{(l+1)}{(2l+1)(2l+3)}\,\sin^{2}\bigl(\delta_{l}^{+}((k)-\delta_{l+1}^{-}(k)\bigr)\Bigr\}.\label{eq:MT}
\end{multline}
Analytic fits of the above momentum-transfer cross-section (and for
other noble gases) are given in McEachran and Stauffer 2014 \cite{McEcStau14}
to aid in plasma modelling calculations.

\begin{figure}[H]
\begin{centering}
\includegraphics[clip,width=0.9\columnwidth]{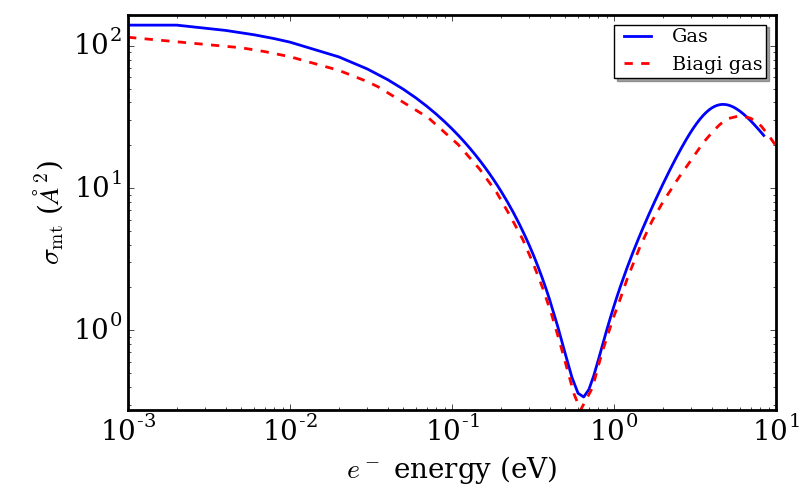}
\par\end{centering}

\raggedright{}\protect\caption{\label{fig:Liquid-Xe-cross-sections-1} The current momentum transfer
cross-sections in the gas-phase (Gas), and the reference momentum
transfer cross-section of Biagi \cite{Bordetal13} (Biagi gas).}
\end{figure}

In Figure \ref{fig:Liquid-Xe-cross-sections-1}, the momentum transfer
cross-section in the gas-phase is compared to the cross-section recommended
by Biagi \cite{lxcatbiagi14a} from the ``Magboltz'' Boltzmann equation
solver. The Biagi elastic momentum transfer cross-section has been
constructed from the unpublished analysis of Elford, fitting to the
available drift velocity and diffusion coefficients \cite{Bordetal13}.
It is often considered the reference cross-section for electron-xenon
interactions in the gas phase. Our momentum transfer cross-section
gives good qualitative agreement with the Biagi reference cross-section,
but generally somewhat over-estimates the value. The location and
depth of the Ramsauer minima, however, agrees closely.

\section{Scattering of electrons by xenon liquid}

\subsection{Xenon pair correlation\label{PairCorrelation}}

The only measurements of the liquid phase xenon structure factor that
are known to us, are by Becchi and Magli \cite{Becchi1997} for $T=274.7$~K
and $N=8.86\times10^{21}\,\mathrm{cm}^{-3}$, near to the critical
point of $T=289.72$~K. In order to obtain a structure factor at
the lower temperature of $T=165$~K and $N=14.2\times10^{22}\,\mathrm{cm}^{-3}$,
where the transport measurements have been performed, we have used
a Lennard-Jones model, calibrated to the high temperature structure
factor, to construct a low temperature structure factor by performing
Monte-Carlo simulations of the fluid.

\begin{figure}[H]
\begin{centering}
\includegraphics[width=0.7\textwidth]{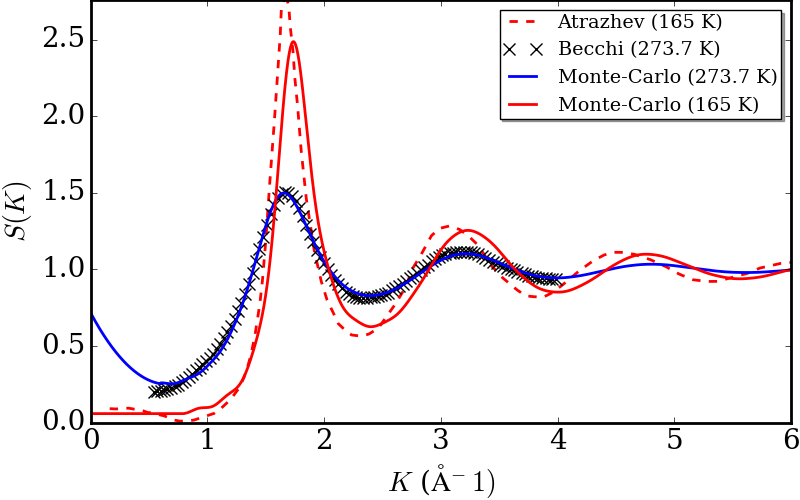}
\par\end{centering}

\protect\caption{\label{fig:structure_factor}Xenon structure factors. The crosses
represent the Becchi measurements \cite{Becchi1997} and the solid
blue lines are our Monte-Carlo simulations of the Lennard-Jones fluid
at a high temperature near the critical point of $274.7$~K. The
solid green line represents the structure factor at $165$~K that
we use in this paper and the dashed green line shows the rescaled
measurements from argon experimental data \cite{Yarnell73} used by
reference \cite{Atrazhevetal05}.}
\end{figure}

The Lennard-Jones fluid has a pair-wise potential given by:
\begin{equation}
V(r)=4\epsilon_{LJ}\left[\left(\frac{r}{\sigma_{LJ}}\right)^{12}-\left(\frac{r}{\sigma_{LJ}}\right)^{6}\right],
\end{equation}
with the characteristic energy $\epsilon_{LJ}$ and length scale $\sigma_{LJ}$.
Often, this potential is truncated to a range of $r_{\mathrm{trunc}}=2.5\,\sigma_{LJ}$
but we work with an effectively untruncated potential by extending
$r_{\mathrm{trunc}}$ to the system size $r_{\mathrm{trunc}}=L/2$.
For the truncated potential, Atrazhev~et~al~\cite{Atrazhevetal05}
have used values of $\epsilon_{LJ}/k_{B}=299\,\mathrm{K}$ and $\sigma_{LJ}=4.05\,\AA$
but, by comparing to the experimental data of reference \cite{Becchi1997},
we find a better fit for our untruncated model by matching the critical
temperature of the Lennard-Jones model ($k_{B}T_{\mathrm{crit}}=1.312\,\epsilon_{LJ}$\ \cite{Perez-Pellitero2006})
to the measured value of $289.72$~K leaving us instead with $\epsilon_{LJ}/k_{B}=220.83$~K.
Given this value, the best fit to the data of reference \cite{Becchi1997}
is $\sigma_{LJ}=3.86\,\AA$. We then use these parameters to obtain
the pair-correlator at the desired temperature of $T=165$~K. A plot
of the various structure factors is shown in figure~\ref{fig:structure_factor},
where we compare with the structure factor calculated from the truncated
potential in the paper of Atrazhev et al. \cite{Atrazhevetal05}.

\subsection{Liquid xenon cross-section\label{sub:Liquid-xenon-cross-section}}

In order to capture the major effects of increasing the number density
of the fluid system, we include two modifications to the gas scattering
potentials due to: a) the screening of the long-range polarization
potential, and b) the influence of the particles in the medium bulk.
The procedure outlined in this section closely follows that of references
\cite{Cohen1967,Boyletal15}. We summarize the steps here. 

Firstly, the effective charge-multipole polarization potential acting
between the electron and the induced multipole of an individual atom
is reduced by the presence of the induced multipoles in the surrounding
atoms, which produces a screening effect. Using the (isotropic) pair-correlator
for xenon, $g(r)$, determined in Section \ref{PairCorrelation},
we have self-consistently calculated the screening function $f(r)$
for an electron located at position $r$ via:

\begin{equation}
f(r)=1-\pi N\int_{0}^{\infty}ds\ \frac{g(s)}{s^{2}}\int_{\left|r-s\right|}^{r+s}dt\ \Theta\left(r,s,t\right)\frac{\alpha(t)f(t)}{t^{2}},
\end{equation}
which has been represented in bipolar coordinates, $s$ and $t$,
and where
\begin{equation}
\Theta\left(r,s,t\right)=\frac{3}{2}\frac{\left(s^{2}+t^{2}-r^{2}\right)\left(r^{2}+t^{2}-s^{2}\right)}{s^{2}}+\left(r^{2}+t^{2}-s^{2}\right),
\end{equation}
arises due to the form of the electric field of a dipole. The screening
function is used to determine the screened polarization potential,
$\tilde{U}_{p}(r),$ of an electron with one atom in a dense fluid, 

\begin{equation}
\tilde{U}_{p}(r)=f(r)U_{p}(r).
\end{equation}

Secondly, the direct interaction of the electrons with other atoms
in the bulk is significant in a dense system, even when the electron
is very close to the focus atom. By following the procedure of Lekner
\cite{Lekner1967}, we construct an effective potential, $U_{\mathrm{eff}}=U_{1}+U_{2}$,
where $U_{1}$ describes the direct (screened) interaction of the
electrons and the target atom, and where $U_{2}$ describes the collective
interaction of the electron with the rest of the bulk atoms. We approximate
$U_{2}$ by an ensemble average of the bulk, i.e., 
\begin{equation}
U_{2}(r)=\frac{2\pi N}{r}\int_{0}^{\infty}dt\ U_{1}(t)\int_{\left|r-t\right|}^{r+t}ds\ sg(s).
\end{equation}
We note that taking the ensemble average has the advantage of enforcing
spherical symmetry of the total effective potential $U_{\mathrm{eff}}$.

Since, in a dense system, the electron is never in effectively free
space, a different measure of the volume ``owned'' by the focus
atom is required. It is natural to define the distance of first turning
point of the potential $U_{\mathrm{eff}}$, which is denoted by $r_{m}$,
as the spherical distance under the influence of the focus atom. Hence
we can say that a single collision event takes place when an electron
enters and leaves the radius $r_{m}$ of a single atom. In order to
calculate the phase shift at the distance $r_{m}$ rather than infinity,
we have set the upper limits of equations (7) and (8) of Boyle et
al. \cite{Boyletal15} for $A_{\kappa}$ and $B_{\kappa}$ to be $r_{m}$.

\begin{figure}[H]
\centering{}\includegraphics[clip,width=0.9\columnwidth]{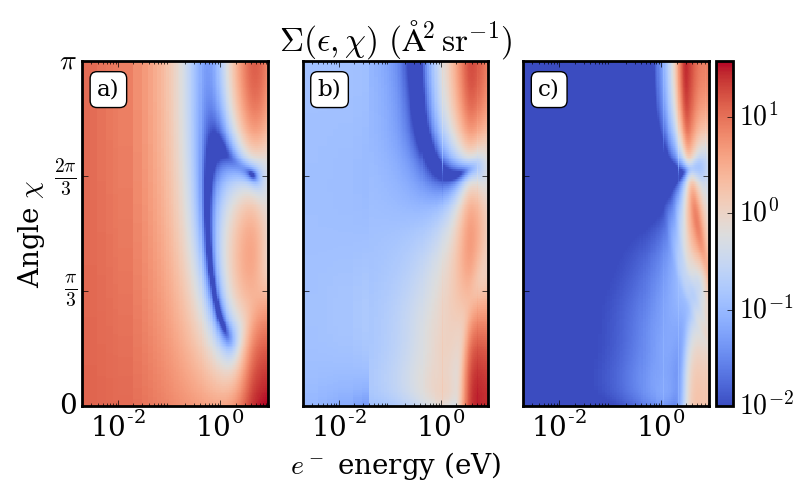}\protect\caption{\label{fig:DifferentialCrossSections} Current differential cross-sections,
$\Sigma(\epsilon,\chi)$., in square angstroms for electrons in Xe
for a) dilute gas phase, b) effective liquid phase including screening
effects, and c) liquid phase including coherent scattering effects.}
\end{figure}

The full differential elastic scattering cross-section for the gas
and liquid phases (with and without coherent scattering) are displayed
in Figure \ref{fig:DifferentialCrossSections}. For the dilute gas
phase, we observe the presence of a forward-peaked minimum in the
range $0.2$--$1$~eV, below which the differential cross-section
is essentially isotropic. This minimimun is the well-known Ramsauer
minimimun which occurs in a number of electron-atom gas-phase cross-sections.
At energies above the minimun, the differential cross-section demonstrates
increased magnitude and also enhanced anisotropy, with peaks in the
forward- and back-scattering directions. When the modification due
to screening and interactions from the liquid bulk are included, we
observe the suppression of the Ramsauer minimum and a removal of the
forward peak for low and moderate energies. At higher energies, the
liquid differential cross-section becomes similar both qualitatively
and quantitatively to the gas phase cross-section. When the liquid
phase differential cross-section is combined with the static structure
factor accounting for coherent scattering effects, the resulting differential
cross-section $\Sigma(\epsilon,\chi)$ takes on a completely different
qualitative structure. At low energies, the differential scattering
cross-section has been further reduced in magnitude at all scattering
angles. At higher energies the forward-scattering has been reduced,
while the back-scattering peak remains essentially unaffected. 

\begin{figure}[H]
\centering{}\includegraphics[clip,width=0.9\columnwidth]{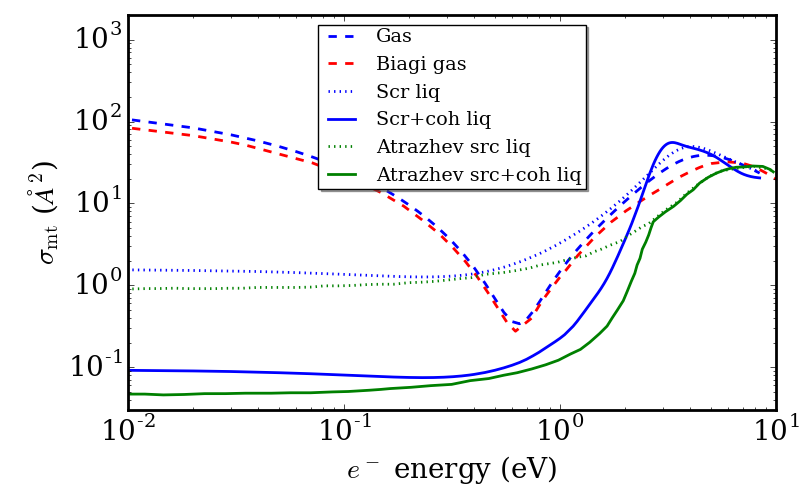}\protect\caption{\label{fig:Liquid-Xe-cross-sections} The current momentum transfer
cross-sections in the gas-phase (Gas; Biagi gas), liquid-phase (Scr
liq, Atrazhev scr liq) and including their modifications when coherent
scattering effects are included (Scr + coh liq, Atrazhev scr + coh
liq). A detailed description of the Atrazhev et al. cross-section
calculations is given in reference \cite{Atrazhevetal05}.}
\end{figure}

The momentum transfer cross-sections corresponding to the differential
scattering cross-sections in Figure \ref{fig:DifferentialCrossSections}
are displayed in Figure \ref{fig:Liquid-Xe-cross-sections}, along
with the cross-sections from Atrazhev et al.~\cite{Atrazhevetal05}.
The comments made regarding the differential scattering cross-sections
are once again reflected here. The Ramsauer minimum observed in the
gas-phase is completely suppressed in the liquid-phase, and there
is a large reduction in the magnitude of the cross-sections when the
screening and liquid effects are included, and again when coherent
scattering effects are included. At higher energies, the liquid-phase
cross-sections approach the gas-phase values, with some additional
oscillatory structure evident when coherent scattering is included.
The cross-sections of Atrazhev et al. \cite{Atrazhevetal05} have
been calculated using a similar formalism, but with a pseudo-potential
that replaces the short-range part of the interaction by a boundary
condition that reproduces the expected scattering length in the gas-phase.
Their cross-sections are qualitatively similar to ours, but are consistently
smaller in magnitude over the range of energies considered. We attribute
these differences to our treatment of the static and exchange parts
of the potential, for both the focus and surrounding atoms, as well
as the inclusion of a full multipole polarization potential.

\section{Results\label{sec:Results}}

One of the key functions of modern-day swarm experiments is to assess
the completeness and accuracy of cross-section \cite{Boyle2014,Boyle2014a,DeUrquijo2014,White2014}
sets. Swarm experiments are many-scattering experiments, where there
is a balance established between the number of particles and the momentum
and energy transfers occurring. In the following sections we consider
the calculation of the macroscopic swarm transport properties in the
gaseous and liquid environments from the microscopic cross-sections.
Particular attention is given to investigating the validity of the
two-term approximation in our calculations, and we compare these with
the full multi-term results. In Section \ref{sub:Electrons-in-liquid-2},
we introduce a simple scaling algorithm to adapt any gas-phase cross-section
to the liquid-phase based on the our \textit{ab initio} scattering
calculations.

\subsection{Electrons in gaseous xenon\label{sub:Electrons-in-gaseous}}

\begin{figure}[H]
\begin{centering}
\includegraphics[width=0.9\columnwidth]{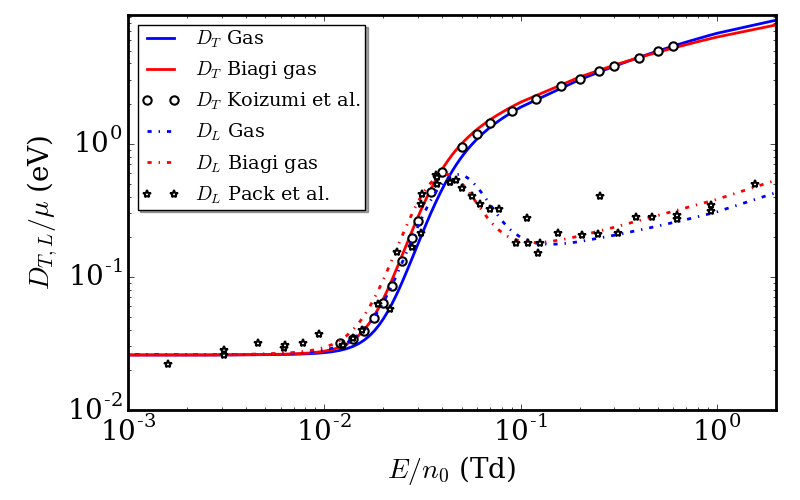}
\par\end{centering}

\begin{centering}
\includegraphics[width=0.9\columnwidth]{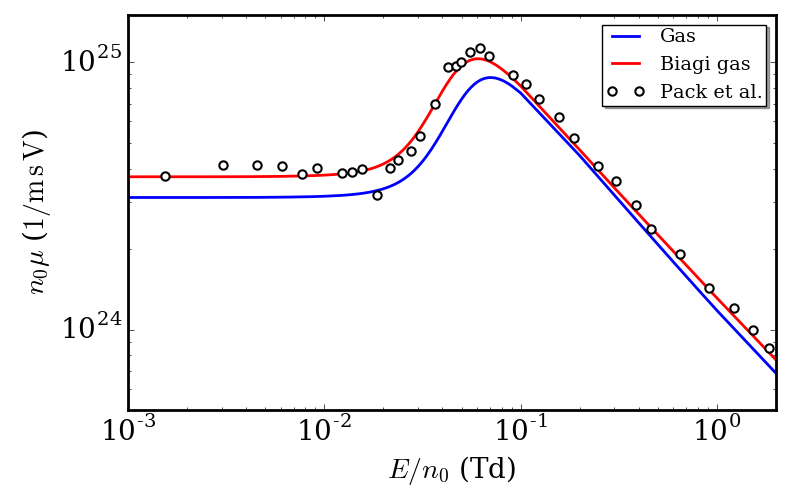}
\par\end{centering}

\protect\caption{\label{fig:Drift-Gas}The transverse ($D_{T}/\mu$) and longitudinal
($D_{L}/\mu$) characteristic energies (top) and reduced mobility
($n_{0}\mu$) (bottom) of electrons in gaseous xenon, calculated using
the current potentials and associated cross-sections detailed in Section
\ref{ScatteringGas} (Gas), and the recommended cross-section of Biagi
\cite{Bordetal13} (Biagi gas), and compared with available experimental
data (Koizumi et al. \cite{koizumi86} at 300~K; Pack et. al \cite{pack92a}
at 300~K;). The background xenon gas for the calculations was fixed
at 300~K.}
\end{figure}

The calculated reduced mobility and characteristic energies using
the gas-phase cross-sections detailed in Section \ref{ScatteringGas}
and the reference cross-sections of Biagi \cite{Bordetal13} are presented
in Figure \ref{fig:Drift-Gas}. They are compared against various
experimental data for xenon gas \cite{koizumi86,pack92a}. We restrict
ourselves to the reduced electric fields of less than $1$ Td, to
ensure we are in the energy regime where only elastic scattering is
operative. For the gas calculations, using the present cross-sections,
we observe agreement to within $30$\% or better for both the reduced
mobility and the transverse characteristic energy over the range of
the reduced fields considered. These errors decrease to $5$\% or
better for the transverse characteristic energy, above the field region
where the transport properties rapidly increase. This can be compared
with the Biagi \cite{Bordetal13} cross-section calculations, which
demonstrate agreement to within $10$\% or better (generally less
than $5$\%) for reduced mobility and transverse characteristic energy.
Although the experimental longitudinal characteristic energies exhibit
larger variation and error than the transverse counterpart, our calculated
energies using the present cross-sections and those of Biagi \cite{Bordetal13}
appear to give good agreement. The major difference between the transport
calculations using the two different cross-sections is in the turning
point of the longitudinal characteristic energy profile: our ab~initio
cross-sections cause a turning point at slightly higher reduced electric
field strengths than the experiment, whereas the Biagi cross-sections
is consistent with the experimental measurements.

In a direct reflection of our increased momentum transfer cross-section,
as compared to the reference, the transverse characteristic energy
and reduced mobility generally underestimate the experimental measurements
\cite{koizumi86,pack92a}. The increased momentum transfer tends to
increase the randomization of the electron's direction during a collision,
which reduces the field\textquoteright s ability to efficiently pump
energy into the system. It is no surprise that the Biagi cross-section
gives closer agreement between calculation and experiment, since it
is informed by experimental swarm measurements and was not from an
\textit{ab initio}\textit{\emph{ theory}}\textit{. }

\subsection{Electrons in liquid xenon\label{sub:Electrons-in-liquid}}

\begin{figure}[H]
\begin{centering}
\includegraphics[width=0.9\columnwidth]{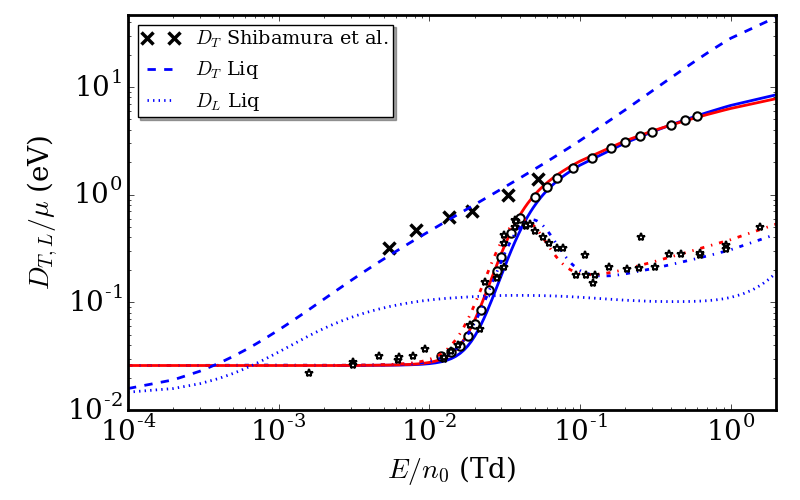}
\par\end{centering}

\begin{centering}
\includegraphics[width=0.9\columnwidth]{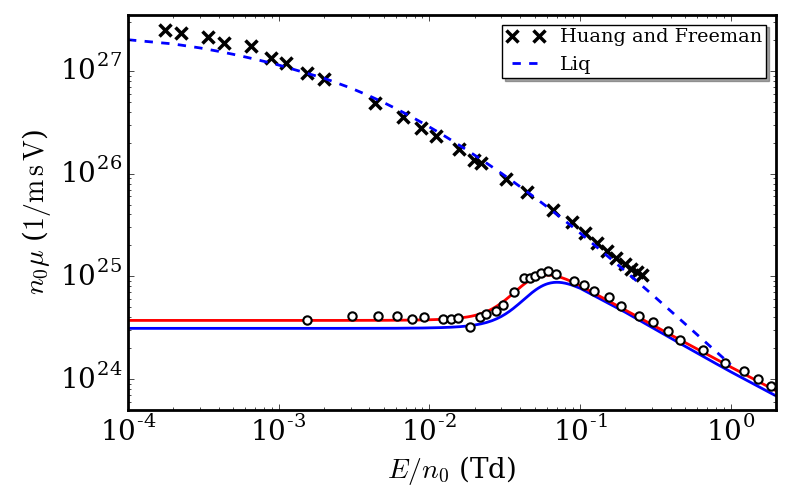}
\par\end{centering}

\protect\caption{\label{fig:Liquid-Ar}Comparison of the transverse ($D_{T}/\mu$)
and longitudinal ($D_{L}/\mu$) characteristic energies  (top) and
reduced mobilities ($n_{0}\mu$) (bottom) in gaseous and liquid xenon,
with those calculated from the various approximations to the cross-sections.
Experimental data ( Liquid phase: Shibamaru et al. \cite{Shibetal1984}
at 165K; Huang and Freeman \cite{HuangFreem78} at 163~K). The current
liquid-phase cross-sections (Liq) have been calculated including coherent
scattering effects. All other profiles are the same as given in Figure
\ref{fig:Drift-Gas}.}
\end{figure}

In Figure \ref{fig:Liquid-Ar} the reduced mobility and characteristic
energies are now compared in both the gaseous and liquid phases. The
transport coefficients are presented against reduced electric fields
so that any linear dependence on density (as occurs in the dilute-gas
limit) has been removed, and so we have a true comparison of the gaseous
and liquid phases. Qualitatively, for a given reduced field in the
low-energy regime, we observe that the reduced mobility and transverse
characteristic energy in the liquid phase are both significantly larger,
often by several orders of magnitude over the gaseous phase. In our
investigations of electrons in liquid argon \cite{Boyletal15}, the
transverse characteristic energy in the liquid phase was instead smaller
than in the gaseous phase. The longitudinal characteristic energy
is generally larger for reduced electric field strengths of less than
$\sim0.2$ Td, but smaller for the higher electric field strength
considered. Our calculations using the ab-initio theory are accurate
to within $25$\% for the mobility (generally less than $15\%$) and
$25$\% for the transverse characteristic energy. These are of similar
magnitude to those for the gas phase. 

\begin{figure}[H]
\centering{}\includegraphics[width=0.9\columnwidth]{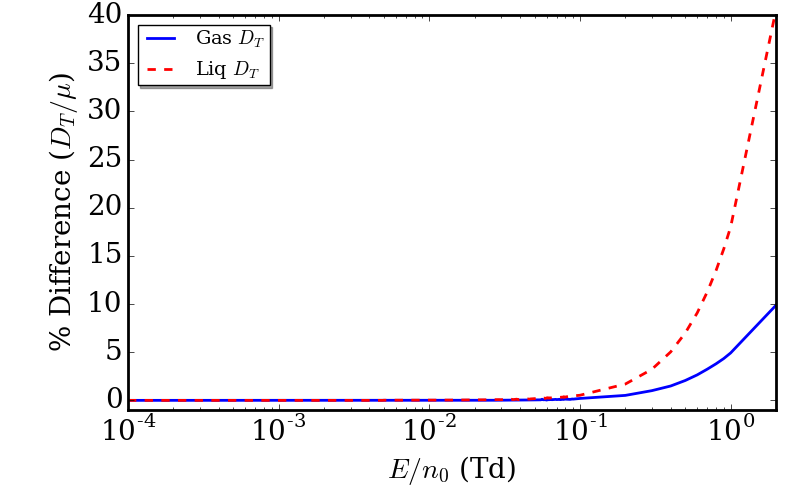}\protect\caption{\label{fig:Two-term approximation} Percentage differences between
the two-term and multi-term values of the characteristic energy for
the gas (Gas) and liquid (Liq) phases. All percentages are relative
to the converged multi-term result using the full differential cross-section. }
\end{figure}

Although the differential scattering cross-sections for Xe displayed
an enhanced anisotropic nature in the liquid phase over that in the
gas phase, the impact of this anisotropy on the velocity distribution
function was not particularly significant here. The errors associated
with the two-term approximation ($l_{\mathrm{max}}=1)$ to the velocity
distribution function are displayed in Figure \ref{fig:Two-term approximation}.
In the gaseous and liquid phases there are differences as large as
$10$\% and $40$\% respectively, in the transverse characteristic
energies. The errors associated with the longitudinal characteristic
energies was less than $0.1$\% for both the gas and liquid phases,
and hence were omitted from Figure \ref{fig:Two-term approximation}.

\subsection{Rescaling gas phase cross-sections \label{sub:Electrons-in-liquid-2}}

\begin{figure}[H]
\centering{}\includegraphics[clip,width=0.9\columnwidth]{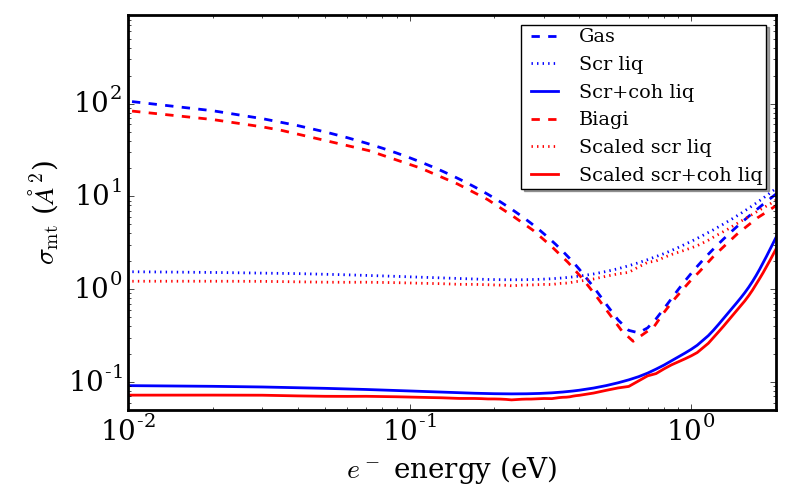}\protect\caption{\label{fig:Liquid-Xe-cross-sections-2} The momentum transfer cross-sections
in the gas-phase (Gas; Biagi), liquid-phase (Scr liq, Scaled scr liq)
and including their modifications when coherent scattering effects
are included ( Scr+coh liq, Scaled scr+coh liq). }
\end{figure}

It was noted in Section \ref{sub:Electrons-in-gaseous} that our gas
phase momentum-transfer cross-section overestimates the reference
cross-section of Biagi \cite{lxcatbiagi14a,Bordetal13}, the latter
of which is shown to produce transport data accurate to within $10$\%
of the experimental data, compared with $30$\% for the current \emph{ab-initio}
theory. In order to utilize the apparent enhanced accuracy associated
with the experimental cross-section, we postulate a method of extracting
the explicit liquid-based effects from the theory and applying it
to the experimentally measured cross-section. The ratio of the liquid
to gas phase cross-section using our formalism gives an energy dependent
scaling for the importance of screening and coherent scattering. Let
us define a scaling factor, $\xi_{l}(v)$, such that:
\begin{equation}
\xi_{l}(v)=\frac{\tilde{\nu_{l}}(v)}{\nu_{l}(v)},
\end{equation}
where these $\nu$ correspond to our scattering calculations using
the Dirac-Fock equations described in Section \ref{ScatteringGas}
and \ref{sub:Liquid-xenon-cross-section} respectively. We can then
build an effective liquid-phase momentum-transfer cross-section based
on the Biagi \cite{Bordetal13} reference data, $\Sigma_{m}^{Biagi}(v)$
, via:
\begin{equation}
\Sigma_{m}^{Biagi}(v)=\xi(v)\sigma_{m}^{Biagi}(v),
\end{equation}
where $\sigma_{m}^{Biagi}(v)$ is the Biagi recommended gas-phase
cross-section \cite{Bordetal13}. It should be noted that, due to
the lack of differential cross-section information for the Biagi set,
only a two-term approximation can be used here, i.e. only the momentum
transfer cross-section is known. The effective Biagi liquid momentum
transfer cross-section is shown in Figure \ref{fig:Liquid-Xe-cross-sections-2}.
The method is similar in spirit to the procedure proposed in reference
\cite{Boyle2012}, though scaling was done with respect to the transport
coefficients rather than the cross-sections. 

\begin{figure}[H]
\begin{centering}
\includegraphics[width=0.9\columnwidth]{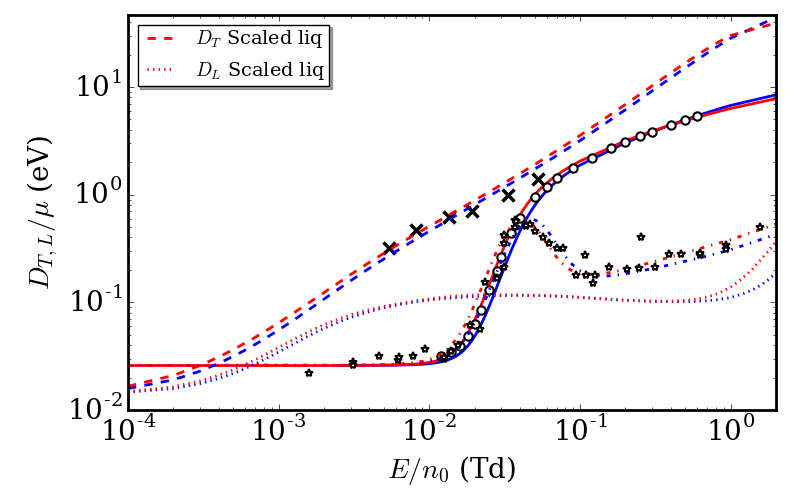}
\par\end{centering}

\begin{centering}
\includegraphics[width=0.9\columnwidth]{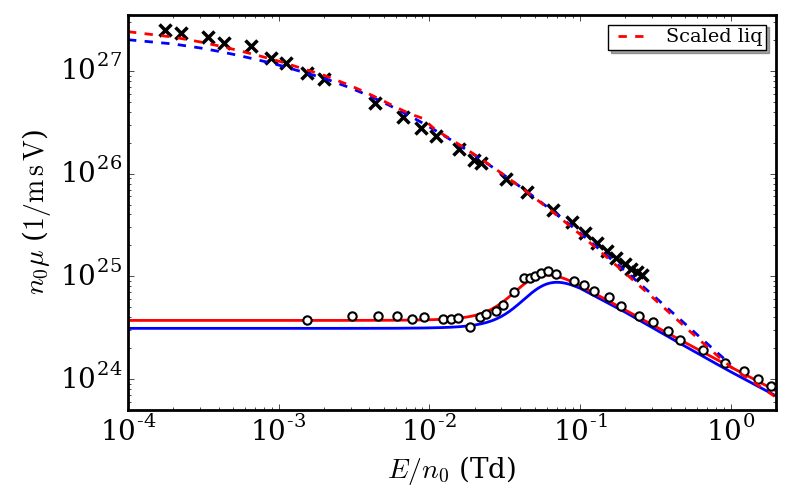}
\par\end{centering}

\protect\caption{\label{fig:Liquid-Xe-1}Comparison of the transverse ($D_{T}/\mu$)
and longitudinal ($D_{L}/\mu$) characteristic energies  (top) and
reduced mobilities ($n_{0}\mu$) (bottom) in gaseous and liquid xenon,
with those calculated from the various approximations to the cross-sections.
The current liquid-phase cross-sections (Liq) have been calculated
including coherent scattering effects. All other profiles are the
same as in Figure \ref{fig:Drift-Gas}. The calculations using the
re-scaled Biagi cross-sections are given by the red  lines (Scaled
liq). All other profiles are the same as given in Figures \ref{fig:Drift-Gas}
and \ref{fig:Liquid-Ar}. }
\end{figure}

The characteristic energies and reduced mobility calculated using
the effective Biagi liquid momentum transfer cross-section is shown
in Figure \ref{fig:Liquid-Xe-1}, along with the previous calculations
considered in this manuscript. It appears that this result gives a
slightly better agreement with experiment for the reduced mobilities
at low field strengths i.e., errors within $20\%$, while it is no
worse for the transverse characteristic energiy. The biggest improvements
occur at the low electric field strengths, where the errors are effectively
half that of the \textit{ab initio} approach.

\section{Conclusions}

We have presented an \emph{ab-initio} treatment of electron scattering
and transport in liquid xenon. There are no free parameters in this
calculation, and hence the agreement to within $25$\% in the mobility
and the transverse characteristic energy in the reduced field range
$10^{-4}$--$1$ Td is considered satisfactory.Given that, in the
dilute gas phase, agreement to within $30$\% for the mobility and
transverse characteristic energy transport coefficients was achieved,
with the current scattering calculations, this gives confidence that
the majority of the essential physics for considering high mobility
noble liquids is present in the theory. A scaling factor formed from
the ratio of the calculated liquid to gas phase cross-sections, was
postulated to encompass the ``liquid''-based effects, enabling more
accurate gas-phase cross-sections derived from experiment to be translated
to liquid-phase cross-sections. Subsequent enhancements in the accuracy
of the transport coefficient calculations were achieved, with the
mobility errors reduced to $20$\% and and even bigger improvement
displayed for the lowest fields considered.
\begin{acknowledgments}
This work was supported in part by the Australian Research Council
through its Discovery Program scheme. M. J. B. and S. J. B. acknowledge
University Malaysia for their ``Icon Professor'' appointments, while
R. D. White thanks Prof. K Ratnavelu for some financial support through
the University Malaysia grant UMRP11A-13AFR. S. D. acknowledges support
from MPNTRRS Projects OI171037 and III41011.
\end{acknowledgments}
\bibliography{library_RW,library,library_extra,ThesisFinal}

\begin{thebibliography}{44}%
\makeatletter
\providecommand \@ifxundefined [1]{%
 \@ifx{#1\undefined}
}%
\providecommand \@ifnum [1]{%
 \ifnum #1\expandafter \@firstoftwo
 \else \expandafter \@secondoftwo
 \fi
}%
\providecommand \@ifx [1]{%
 \ifx #1\expandafter \@firstoftwo
 \else \expandafter \@secondoftwo
 \fi
}%
\providecommand \natexlab [1]{#1}%
\providecommand \enquote  [1]{``#1''}%
\providecommand \bibnamefont  [1]{#1}%
\providecommand \bibfnamefont [1]{#1}%
\providecommand \citenamefont [1]{#1}%
\providecommand \href@noop [0]{\@secondoftwo}%
\providecommand \href [0]{\begingroup \@sanitize@url \@href}%
\providecommand \@href[1]{\@@startlink{#1}\@@href}%
\providecommand \@@href[1]{\endgroup#1\@@endlink}%
\providecommand \@sanitize@url [0]{\catcode `\\12\catcode `\$12\catcode
  `\&12\catcode `\#12\catcode `\^12\catcode `\_12\catcode `\%12\relax}%
\providecommand \@@startlink[1]{}%
\providecommand \@@endlink[0]{}%
\providecommand \url  [0]{\begingroup\@sanitize@url \@url }%
\providecommand \@url [1]{\endgroup\@href {#1}{\urlprefix }}%
\providecommand \urlprefix  [0]{URL }%
\providecommand \Eprint [0]{\href }%
\providecommand \doibase [0]{http://dx.doi.org/}%
\providecommand \selectlanguage [0]{\@gobble}%
\providecommand \bibinfo  [0]{\@secondoftwo}%
\providecommand \bibfield  [0]{\@secondoftwo}%
\providecommand \translation [1]{[#1]}%
\providecommand \BibitemOpen [0]{}%
\providecommand \bibitemStop [0]{}%
\providecommand \bibitemNoStop [0]{.\EOS\space}%
\providecommand \EOS [0]{\spacefactor3000\relax}%
\providecommand \BibitemShut  [1]{\csname bibitem#1\endcsname}%
\let\auto@bib@innerbib\@empty
\bibitem [{\citenamefont {Boyle}\ \emph {et~al.}(2015)\citenamefont {Boyle},
  \citenamefont {McEachran}, \citenamefont {Cocks},\ and\ \citenamefont
  {White}}]{Boyletal15}%
  \BibitemOpen
  \bibfield  {author} {\bibinfo {author} {\bibfnamefont {G.~J.}\ \bibnamefont
  {Boyle}}, \bibinfo {author} {\bibfnamefont {R.~P.}\ \bibnamefont
  {McEachran}}, \bibinfo {author} {\bibfnamefont {D.~G.}\ \bibnamefont
  {Cocks}}, \ and\ \bibinfo {author} {\bibfnamefont {R.~D.}\ \bibnamefont
  {White}},\ }\href@noop {} {\bibfield  {journal} {\bibinfo  {journal} {J.
  Chem. Phys.}\ }\textbf {\bibinfo {volume} {142}},\ \bibinfo {pages} {154507}
  (\bibinfo {year} {2015})}\BibitemShut {NoStop}%
\bibitem [{\citenamefont {{C. Regenfus and the ArDM
  collaboration}}(2010)}]{Regenfus2010}%
  \BibitemOpen
  \bibfield  {author} {\bibinfo {author} {\bibnamefont {{C. Regenfus and the
  ArDM collaboration}}},\ }\href@noop {} {\bibfield  {journal} {\bibinfo
  {journal} {J. Phys. Conf. Ser.}\ }\textbf {\bibinfo {volume} {203}},\
  \bibinfo {pages} {012024} (\bibinfo {year} {2010})}\BibitemShut {NoStop}%
\bibitem [{\citenamefont {Aprile}\ and\ \citenamefont
  {Doke}(2010)}]{Aprile2010}%
  \BibitemOpen
  \bibfield  {author} {\bibinfo {author} {\bibfnamefont {E.}~\bibnamefont
  {Aprile}}\ and\ \bibinfo {author} {\bibfnamefont {T.}~\bibnamefont {Doke}},\
  }\href {\doibase 10.1103/RevModPhys.82.2053} {\bibfield  {journal} {\bibinfo
  {journal} {Rev. Mod. Phys.}\ }\textbf {\bibinfo {volume} {82}},\ \bibinfo
  {pages} {2053} (\bibinfo {year} {2010})}\BibitemShut {NoStop}%
\bibitem [{\citenamefont {{E. Aprile, J. Angle, F. Arneodo, L. Baudis, A.
  Bernstein, A. Bolozdynya, et al}}(2011)}]{Aprile2011g}%
  \BibitemOpen
  \bibfield  {author} {\bibinfo {author} {\bibnamefont {{E. Aprile, J. Angle,
  F. Arneodo, L. Baudis, A. Bernstein, A. Bolozdynya, et al}}},\ }\href@noop {}
  {\bibfield  {journal} {\bibinfo  {journal} {Astropart. Phys.}\ }\textbf
  {\bibinfo {volume} {34}},\ \bibinfo {pages} {679} (\bibinfo {year}
  {2011})}\BibitemShut {NoStop}%
\bibitem [{\citenamefont {{G. J. Alner, H. M. Ara{\'u}jo, A. Bewick, C. Bungau,
  B. Camanzi, M. J. Carson, et al}}(2007)}]{Alner2007g}%
  \BibitemOpen
  \bibfield  {author} {\bibinfo {author} {\bibnamefont {{G. J. Alner, H. M.
  Ara{\'u}jo, A. Bewick, C. Bungau, B. Camanzi, M. J. Carson, et al}}},\
  }\href@noop {} {\bibfield  {journal} {\bibinfo  {journal} {Astropart. Phys.}\
  }\textbf {\bibinfo {volume} {28}},\ \bibinfo {pages} {287} (\bibinfo {year}
  {2007})}\BibitemShut {NoStop}%
\bibitem [{\citenamefont {{B. Baller, C. Bromberg, N. Buchanan, F. Cavanna, H.
  Chen, E. Church, et al}}(2014)}]{baller2014liquid}%
  \BibitemOpen
  \bibfield  {author} {\bibinfo {author} {\bibnamefont {{B. Baller, C.
  Bromberg, N. Buchanan, F. Cavanna, H. Chen, E. Church, et al}}},\ }\href@noop
  {} {\bibfield  {journal} {\bibinfo  {journal} {J. Instrum.}\ }\textbf
  {\bibinfo {volume} {9}},\ \bibinfo {pages} {T05005} (\bibinfo {year}
  {2014})}\BibitemShut {NoStop}%
\bibitem [{\citenamefont {{S. Amerio, S. Amoruso, M. Antonello, P. Aprili, M.
  Armenante, F. Arneodo, et al}}(2004)}]{ICARUS2004}%
  \BibitemOpen
  \bibfield  {author} {\bibinfo {author} {\bibnamefont {{S. Amerio, S. Amoruso,
  M. Antonello, P. Aprili, M. Armenante, F. Arneodo, et al}}},\ }\href@noop {}
  {\bibfield  {journal} {\bibinfo  {journal} {Nucl. Instr. Meth. Phys. Res. A}\
  }\textbf {\bibinfo {volume} {527}},\ \bibinfo {pages} {329} (\bibinfo {year}
  {2004})}\BibitemShut {NoStop}%
\bibitem [{\citenamefont {{C. Rubbia, M. Antonello, P. Aprili, B. Baibussinov,
  M. Baldo Ceolin, L. Barz\'e, et al}}(2011)}]{ICARUS2011}%
  \BibitemOpen
  \bibfield  {author} {\bibinfo {author} {\bibnamefont {{C. Rubbia, M.
  Antonello, P. Aprili, B. Baibussinov, M. Baldo Ceolin, L. Barz\'e, et al}}},\
  }\href@noop {} {\bibfield  {journal} {\bibinfo  {journal} {J. Instrum.}\
  }\textbf {\bibinfo {volume} {6}},\ \bibinfo {pages} {P07011} (\bibinfo {year}
  {2011})}\BibitemShut {NoStop}%
\bibitem [{\citenamefont {{The ATLAS collaboration}}(2014)}]{ATLAS2014}%
  \BibitemOpen
  \bibfield  {author} {\bibinfo {author} {\bibnamefont {{The ATLAS
  collaboration}}},\ }\href@noop {} {\bibfield  {journal} {\bibinfo  {journal}
  {J. Instrum.}\ }\textbf {\bibinfo {volume} {9}},\ \bibinfo {pages} {P07024}
  (\bibinfo {year} {2014})}\BibitemShut {NoStop}%
\bibitem [{\citenamefont {{Gee}}\ and\ \citenamefont
  {{Freeman}}(1986)}]{GeeFree86}%
  \BibitemOpen
  \bibfield  {author} {\bibinfo {author} {\bibfnamefont {N.}~\bibnamefont
  {{Gee}}}\ and\ \bibinfo {author} {\bibfnamefont {G.~R.}\ \bibnamefont
  {{Freeman}}},\ }\href@noop {} {\bibfield  {journal} {\bibinfo  {journal}
  {Can. J. Chem.}\ }\textbf {\bibinfo {volume} {64}},\ \bibinfo {pages} {1810}
  (\bibinfo {year} {1986})}\BibitemShut {NoStop}%
\bibitem [{\citenamefont {Braglia}\ and\ \citenamefont
  {Dallacasa}(1982)}]{Braglia82}%
  \BibitemOpen
  \bibfield  {author} {\bibinfo {author} {\bibfnamefont {G.}~\bibnamefont
  {Braglia}}\ and\ \bibinfo {author} {\bibfnamefont {V.}~\bibnamefont
  {Dallacasa}},\ }\href@noop {} {\bibfield  {journal} {\bibinfo  {journal}
  {Phys. Rev. A}\ }\textbf {\bibinfo {volume} {26}},\ \bibinfo {pages} {902}
  (\bibinfo {year} {1982})}\BibitemShut {NoStop}%
\bibitem [{\citenamefont {Borghesani}(2006)}]{Borghesani2006}%
  \BibitemOpen
  \bibfield  {author} {\bibinfo {author} {\bibfnamefont {A.~F.}\ \bibnamefont
  {Borghesani}},\ }\href {\doibase 10.1109/TDEI.2006.1657960} {\bibfield
  {journal} {\bibinfo  {journal} {IEEE Transactions on Dielectrics and
  Electrical Insulation}\ }\textbf {\bibinfo {volume} {13}},\ \bibinfo {pages}
  {492} (\bibinfo {year} {2006})}\BibitemShut {NoStop}%
\bibitem [{\citenamefont {Sakai}(2007{\natexlab{a}})}]{Sakai2007}%
  \BibitemOpen
  \bibfield  {author} {\bibinfo {author} {\bibfnamefont {Y.}~\bibnamefont
  {Sakai}},\ }\href {\doibase 10.1088/0022-3727/40/24/R01} {\bibfield
  {journal} {\bibinfo  {journal} {Journal of Physics D: Applied Physics}\
  }\textbf {\bibinfo {volume} {40}},\ \bibinfo {pages} {R441} (\bibinfo {year}
  {2007}{\natexlab{a}})}\BibitemShut {NoStop}%
\bibitem [{\citenamefont {White}\ and\ \citenamefont
  {Robson}(2009)}]{White2009}%
  \BibitemOpen
  \bibfield  {author} {\bibinfo {author} {\bibfnamefont {R.~D.}\ \bibnamefont
  {White}}\ and\ \bibinfo {author} {\bibfnamefont {R.~E.}\ \bibnamefont
  {Robson}},\ }\href {http://link.aps.org/doi/10.1103/PhysRevLett.102.230602}
  {\bibfield  {journal} {\bibinfo  {journal} {Phys. Rev. Lett.}\ }\textbf
  {\bibinfo {volume} {102}},\ \bibinfo {pages} {230602} (\bibinfo {year}
  {2009})}\BibitemShut {NoStop}%
\bibitem [{\citenamefont {Kunhardt}(1991)}]{Kunhardt1991}%
  \BibitemOpen
  \bibfield  {author} {\bibinfo {author} {\bibfnamefont {E.~E.}\ \bibnamefont
  {Kunhardt}},\ }\href {\doibase 10.1103/PhysRevB.44.4235} {\bibfield
  {journal} {\bibinfo  {journal} {Physical Review B}\ }\textbf {\bibinfo
  {volume} {44}},\ \bibinfo {pages} {4235} (\bibinfo {year}
  {1991})}\BibitemShut {NoStop}%
\bibitem [{\citenamefont {Cohen}\ and\ \citenamefont
  {Lekner}(1967)}]{Cohen1967}%
  \BibitemOpen
  \bibfield  {author} {\bibinfo {author} {\bibfnamefont {M.~H.}\ \bibnamefont
  {Cohen}}\ and\ \bibinfo {author} {\bibfnamefont {J.}~\bibnamefont {Lekner}},\
  }\href@noop {} {\bibfield  {journal} {\bibinfo  {journal} {Physical Review}\
  }\textbf {\bibinfo {volume} {158}},\ \bibinfo {pages} {305} (\bibinfo {year}
  {1967})}\BibitemShut {NoStop}%
\bibitem [{\citenamefont {Atrazhev}\ and\ \citenamefont
  {Iakubov}(1981)}]{Atrazhev1981}%
  \BibitemOpen
  \bibfield  {author} {\bibinfo {author} {\bibfnamefont {V.~M.}\ \bibnamefont
  {Atrazhev}}\ and\ \bibinfo {author} {\bibfnamefont {I.~T.}\ \bibnamefont
  {Iakubov}},\ }\href {\doibase 10.1088/0022-3719/14/33/021} {\bibfield
  {journal} {\bibinfo  {journal} {Journal of Physics C: Solid State Physics}\
  }\textbf {\bibinfo {volume} {14}},\ \bibinfo {pages} {5139} (\bibinfo {year}
  {1981})}\BibitemShut {NoStop}%
\bibitem [{\citenamefont {Atrazhev}\ and\ \citenamefont
  {Timoshkin}(1996)}]{Atrazhev1996}%
  \BibitemOpen
  \bibfield  {author} {\bibinfo {author} {\bibfnamefont {V.}~\bibnamefont
  {Atrazhev}}\ and\ \bibinfo {author} {\bibfnamefont {I.}~\bibnamefont
  {Timoshkin}},\ }\href {\doibase 10.1103/PhysRevB.54.11252} {\bibfield
  {journal} {\bibinfo  {journal} {Physical Review B}\ }\textbf {\bibinfo
  {volume} {54}},\ \bibinfo {pages} {11252} (\bibinfo {year}
  {1996})}\BibitemShut {NoStop}%
\bibitem [{\citenamefont {White}\ and\ \citenamefont
  {Robson}(2011)}]{White2011}%
  \BibitemOpen
  \bibfield  {author} {\bibinfo {author} {\bibfnamefont {R.~D.}\ \bibnamefont
  {White}}\ and\ \bibinfo {author} {\bibfnamefont {R.~E.}\ \bibnamefont
  {Robson}},\ }\href@noop {} {\bibfield  {journal} {\bibinfo  {journal}
  {Physical Review E}\ }\textbf {\bibinfo {volume} {84}},\ \bibinfo {pages}
  {031125} (\bibinfo {year} {2011})}\BibitemShut {NoStop}%
\bibitem [{\citenamefont {White}\ \emph {et~al.}(2003)\citenamefont {White},
  \citenamefont {Robson}, \citenamefont {Schmidt},\ and\ \citenamefont
  {Morrison}}]{White2003a}%
  \BibitemOpen
  \bibfield  {author} {\bibinfo {author} {\bibfnamefont {R.~D.}\ \bibnamefont
  {White}}, \bibinfo {author} {\bibfnamefont {R.~E.}\ \bibnamefont {Robson}},
  \bibinfo {author} {\bibfnamefont {B.}~\bibnamefont {Schmidt}}, \ and\
  \bibinfo {author} {\bibfnamefont {M.}~\bibnamefont {Morrison}},\ }\href@noop
  {} {\bibfield  {journal} {\bibinfo  {journal} {Journal of Physics D: Applied
  Physics}\ }\textbf {\bibinfo {volume} {36}},\ \bibinfo {pages} {3125}
  (\bibinfo {year} {2003})}\BibitemShut {NoStop}%
\bibitem [{\citenamefont {Boltzmann}(1872)}]{Boltzmann1872}%
  \BibitemOpen
  \bibfield  {author} {\bibinfo {author} {\bibfnamefont {L.}~\bibnamefont
  {Boltzmann}},\ }\href@noop {} {\bibfield  {journal} {\bibinfo  {journal}
  {Wein. Ber.}\ }\textbf {\bibinfo {volume} {66}},\ \bibinfo {pages} {275}
  (\bibinfo {year} {1872})}\BibitemShut {NoStop}%
\bibitem [{\citenamefont {Huxley}\ and\ \citenamefont
  {Crompton}(1974)}]{HuxlCrom74}%
  \BibitemOpen
  \bibfield  {author} {\bibinfo {author} {\bibfnamefont {L.~G.~H.}\
  \bibnamefont {Huxley}}\ and\ \bibinfo {author} {\bibfnamefont {R.~W.}\
  \bibnamefont {Crompton}},\ }\href@noop {} {\emph {\bibinfo {title} {The Drift
  and Diffusion of Electrons in Gases}}}\ (\bibinfo  {publisher} {Wiley},\
  \bibinfo {address} {New York},\ \bibinfo {year} {1974})\BibitemShut {NoStop}%
\bibitem [{\citenamefont {Sakai}(2007{\natexlab{b}})}]{Sakai2007a}%
  \BibitemOpen
  \bibfield  {author} {\bibinfo {author} {\bibfnamefont {Y.}~\bibnamefont
  {Sakai}},\ }\href {\doibase 10.1088/0022-3727/40/24/R01} {\bibfield
  {journal} {\bibinfo  {journal} {Journal of Physics D: Applied Physics}\
  }\textbf {\bibinfo {volume} {40}},\ \bibinfo {pages} {R441} (\bibinfo {year}
  {2007}{\natexlab{b}})}\BibitemShut {NoStop}%
\bibitem [{\citenamefont {Chen}\ \emph {et~al.}(2008)\citenamefont {Chen},
  \citenamefont {McEachran},\ and\ \citenamefont {Stauffer}}]{Chen08}%
  \BibitemOpen
  \bibfield  {author} {\bibinfo {author} {\bibfnamefont {S.}~\bibnamefont
  {Chen}}, \bibinfo {author} {\bibfnamefont {R.~P.}\ \bibnamefont {McEachran}},
  \ and\ \bibinfo {author} {\bibfnamefont {A.~D.}\ \bibnamefont {Stauffer}},\
  }\href@noop {} {\bibfield  {journal} {\bibinfo  {journal} {J. Phys. B}\
  }\textbf {\bibinfo {volume} {41}},\ \bibinfo {pages} {025201} (\bibinfo
  {year} {2008})}\BibitemShut {NoStop}%
\bibitem [{\citenamefont {McEachran}\ \emph {et~al.}(1977)\citenamefont
  {McEachran}, \citenamefont {Stauffer}, \citenamefont {Ryman},\ and\
  \citenamefont {Morgan}}]{mceachran77}%
  \BibitemOpen
  \bibfield  {author} {\bibinfo {author} {\bibfnamefont {R.~P.}\ \bibnamefont
  {McEachran}}, \bibinfo {author} {\bibfnamefont {A.~D.}\ \bibnamefont
  {Stauffer}}, \bibinfo {author} {\bibfnamefont {A.~G.}\ \bibnamefont {Ryman}},
  \ and\ \bibinfo {author} {\bibfnamefont {D.~L.}\ \bibnamefont {Morgan}},\
  }\href@noop {} {\bibfield  {journal} {\bibinfo  {journal} {J. Phys. B}\
  }\textbf {\bibinfo {volume} {10}},\ \bibinfo {pages} {663} (\bibinfo {year}
  {1977})}\BibitemShut {NoStop}%
\bibitem [{\citenamefont {McEachran}\ and\ \citenamefont
  {Stauffer}(1990)}]{McEachran90}%
  \BibitemOpen
  \bibfield  {author} {\bibinfo {author} {\bibfnamefont {R.~P.}\ \bibnamefont
  {McEachran}}\ and\ \bibinfo {author} {\bibfnamefont {A.~D.}\ \bibnamefont
  {Stauffer}},\ }\href@noop {} {\bibfield  {journal} {\bibinfo  {journal} {J.
  Phys. B}\ }\textbf {\bibinfo {volume} {23}},\ \bibinfo {pages} {4605}
  (\bibinfo {year} {1990})}\BibitemShut {NoStop}%
\bibitem [{\citenamefont {Mimnagh}\ \emph {et~al.}(1993)\citenamefont
  {Mimnagh}, \citenamefont {McEachran},\ and\ \citenamefont
  {Stauffer}}]{Mimnagh93}%
  \BibitemOpen
  \bibfield  {author} {\bibinfo {author} {\bibfnamefont {D.~J.~R.}\
  \bibnamefont {Mimnagh}}, \bibinfo {author} {\bibfnamefont {R.~P.}\
  \bibnamefont {McEachran}}, \ and\ \bibinfo {author} {\bibfnamefont {A.~D.}\
  \bibnamefont {Stauffer}},\ }\href@noop {} {\bibfield  {journal} {\bibinfo
  {journal} {J. Phys. B}\ }\textbf {\bibinfo {volume} {26}},\ \bibinfo {pages}
  {1727} (\bibinfo {year} {1993})}\BibitemShut {NoStop}%
\bibitem [{\citenamefont {{McEachran}}\ and\ \citenamefont
  {{Stauffer}}(2014)}]{McEcStau14}%
  \BibitemOpen
  \bibfield  {author} {\bibinfo {author} {\bibfnamefont {R.~P.}\ \bibnamefont
  {{McEachran}}}\ and\ \bibinfo {author} {\bibfnamefont {A.~D.}\ \bibnamefont
  {{Stauffer}}},\ }\href@noop {} {\bibfield  {journal} {\bibinfo  {journal}
  {Eurl. Phys. J.}\ }\textbf {\bibinfo {volume} {68}},\ \bibinfo {pages} {153}
  (\bibinfo {year} {2014})}\BibitemShut {NoStop}%
\bibitem [{\citenamefont {{Biagi}}(2014)}]{lxcatbiagi14a}%
  \BibitemOpen
  \bibfield  {author} {\bibinfo {author} {\bibfnamefont {S.~F.}\ \bibnamefont
  {{Biagi}}},\ }\href {http://www.lxcat.net/} {\enquote {\bibinfo {title}
  {{Biagi database}},}\ }\bibinfo {howpublished} {{www.lxcat.net}} (\bibinfo
  {year} {2014}),\ \bibinfo {note} {accessed 30 April, 2014}\BibitemShut
  {NoStop}%
\bibitem [{\citenamefont {{Bordage}}\ \emph {et~al.}(2013)\citenamefont
  {{Bordage}}, \citenamefont {{Biagi}}, \citenamefont {{Alves}}, \citenamefont
  {{Bartschat}}, \citenamefont {{Chowdhury}}, \citenamefont {{Pitchford}},
  \citenamefont {{Hagelaar}}, \citenamefont {{Morgan}},\ and\ \citenamefont
  {{Zatsarinny}}}]{Bordetal13}%
  \BibitemOpen
  \bibfield  {author} {\bibinfo {author} {\bibfnamefont {M.~C.}\ \bibnamefont
  {{Bordage}}}, \bibinfo {author} {\bibfnamefont {S.~F.}\ \bibnamefont
  {{Biagi}}}, \bibinfo {author} {\bibfnamefont {L.~L.}\ \bibnamefont
  {{Alves}}}, \bibinfo {author} {\bibfnamefont {K.}~\bibnamefont
  {{Bartschat}}}, \bibinfo {author} {\bibfnamefont {S.}~\bibnamefont
  {{Chowdhury}}}, \bibinfo {author} {\bibfnamefont {L.~C.}\ \bibnamefont
  {{Pitchford}}}, \bibinfo {author} {\bibfnamefont {G.~J.~M.}\ \bibnamefont
  {{Hagelaar}}}, \bibinfo {author} {\bibfnamefont {V.~P.}\ \bibnamefont
  {{Morgan}}}, \ and\ \bibinfo {author} {\bibfnamefont {O.}~\bibnamefont
  {{Zatsarinny}}},\ }\href@noop {} {\bibfield  {journal} {\bibinfo  {journal}
  {J. Phys. D: Appl. Phys.}\ }\textbf {\bibinfo {volume} {46}},\ \bibinfo
  {pages} {334003} (\bibinfo {year} {2013})}\BibitemShut {NoStop}%
\bibitem [{\citenamefont {Becchi}\ and\ \citenamefont
  {Magli}(1997)}]{Becchi1997}%
  \BibitemOpen
  \bibfield  {author} {\bibinfo {author} {\bibfnamefont {M.}~\bibnamefont
  {Becchi}}\ and\ \bibinfo {author} {\bibfnamefont {R.}~\bibnamefont {Magli}},\
  }\href {\doibase 10.1016/S0921-4526(96)00970-2} {\bibfield  {journal}
  {\bibinfo  {journal} {Physica B: Condensed Matter}\ }\textbf {\bibinfo
  {volume} {234}},\ \bibinfo {pages} {316} (\bibinfo {year}
  {1997})}\BibitemShut {NoStop}%
\bibitem [{\citenamefont {{Atrazhev}}\ \emph {et~al.}(2005)\citenamefont
  {{Atrazhev}}, \citenamefont {{Berezhnov}}, \citenamefont {{Dunikov}},\ and\
  \citenamefont {{Chernysheva}}}]{Atrazhevetal05}%
  \BibitemOpen
  \bibfield  {author} {\bibinfo {author} {\bibfnamefont {V.~M.}\ \bibnamefont
  {{Atrazhev}}}, \bibinfo {author} {\bibfnamefont {A.~V.}\ \bibnamefont
  {{Berezhnov}}}, \bibinfo {author} {\bibfnamefont {D.~O.}\ \bibnamefont
  {{Dunikov}}}, \ and\ \bibinfo {author} {\bibfnamefont {I.~V.}\ \bibnamefont
  {{Chernysheva}}},\ }in\ \href@noop {} {\emph {\bibinfo {booktitle} {IEEE
  International Conference on Dielectric Liquids}}}\ (\bibinfo {year} {2005})\
  pp.\ \bibinfo {pages} {329--332}\BibitemShut {NoStop}%
\bibitem [{\citenamefont {P{\'{e}}rez-Pellitero}\ \emph
  {et~al.}(2006)\citenamefont {P{\'{e}}rez-Pellitero}, \citenamefont {Ungerer},
  \citenamefont {Orkoulas},\ and\ \citenamefont
  {Mackie}}]{Perez-Pellitero2006}%
  \BibitemOpen
  \bibfield  {author} {\bibinfo {author} {\bibfnamefont {J.}~\bibnamefont
  {P{\'{e}}rez-Pellitero}}, \bibinfo {author} {\bibfnamefont {P.}~\bibnamefont
  {Ungerer}}, \bibinfo {author} {\bibfnamefont {G.}~\bibnamefont {Orkoulas}}, \
  and\ \bibinfo {author} {\bibfnamefont {A.~D.}\ \bibnamefont {Mackie}},\
  }\href {\doibase 10.1063/1.2227027} {\bibfield  {journal} {\bibinfo
  {journal} {Journal of Chemical Physics}\ }\textbf {\bibinfo {volume} {125}},\
  \bibinfo {pages} {1} (\bibinfo {year} {2006})}\BibitemShut {NoStop}%
\bibitem [{\citenamefont {Lekner}(1967)}]{Lekner1967}%
  \BibitemOpen
  \bibfield  {author} {\bibinfo {author} {\bibfnamefont {J.}~\bibnamefont
  {Lekner}},\ }\href@noop {} {\bibfield  {journal} {\bibinfo  {journal}
  {Physical Review}\ }\textbf {\bibinfo {volume} {158}},\ \bibinfo {pages}
  {103} (\bibinfo {year} {1967})}\BibitemShut {NoStop}%
\bibitem [{\citenamefont {Boyle}\ \emph
  {et~al.}(2014{\natexlab{a}})\citenamefont {Boyle}, \citenamefont {Casey},
  \citenamefont {White},\ and\ \citenamefont {Mitroy}}]{Boyle2014}%
  \BibitemOpen
  \bibfield  {author} {\bibinfo {author} {\bibfnamefont {G.}~\bibnamefont
  {Boyle}}, \bibinfo {author} {\bibfnamefont {M.}~\bibnamefont {Casey}},
  \bibinfo {author} {\bibfnamefont {R.}~\bibnamefont {White}}, \ and\ \bibinfo
  {author} {\bibfnamefont {J.}~\bibnamefont {Mitroy}},\ }\href {\doibase
  10.1103/PhysRevA.89.022712} {\bibfield  {journal} {\bibinfo  {journal}
  {Physical Review A}\ }\textbf {\bibinfo {volume} {89}},\ \bibinfo {pages}
  {022712} (\bibinfo {year} {2014}{\natexlab{a}})}\BibitemShut {NoStop}%
\bibitem [{\citenamefont {Boyle}\ \emph
  {et~al.}(2014{\natexlab{b}})\citenamefont {Boyle}, \citenamefont {Casey},
  \citenamefont {White}, \citenamefont {Cheng},\ and\ \citenamefont
  {Mitroy}}]{Boyle2014a}%
  \BibitemOpen
  \bibfield  {author} {\bibinfo {author} {\bibfnamefont {G.~J.}\ \bibnamefont
  {Boyle}}, \bibinfo {author} {\bibfnamefont {M.~J.~E.}\ \bibnamefont {Casey}},
  \bibinfo {author} {\bibfnamefont {R.~D.}\ \bibnamefont {White}}, \bibinfo
  {author} {\bibfnamefont {Y.}~\bibnamefont {Cheng}}, \ and\ \bibinfo {author}
  {\bibfnamefont {J.}~\bibnamefont {Mitroy}},\ }\href {\doibase
  10.1088/0022-3727/47/34/345203} {\bibfield  {journal} {\bibinfo  {journal}
  {Journal of Physics D: Applied Physics}\ }\textbf {\bibinfo {volume} {47}},\
  \bibinfo {pages} {345203} (\bibinfo {year} {2014}{\natexlab{b}})}\BibitemShut
  {NoStop}%
\bibitem [{\citenamefont {de~Urquijo}\ \emph {et~al.}(2014)\citenamefont
  {de~Urquijo}, \citenamefont {Basurto}, \citenamefont {Ju{\'{a}}rez},
  \citenamefont {Ness}, \citenamefont {Robson}, \citenamefont {Brunger},\ and\
  \citenamefont {White}}]{DeUrquijo2014}%
  \BibitemOpen
  \bibfield  {author} {\bibinfo {author} {\bibfnamefont {J.}~\bibnamefont
  {de~Urquijo}}, \bibinfo {author} {\bibfnamefont {E.}~\bibnamefont {Basurto}},
  \bibinfo {author} {\bibfnamefont {a.~M.}\ \bibnamefont {Ju{\'{a}}rez}},
  \bibinfo {author} {\bibfnamefont {K.~F.}\ \bibnamefont {Ness}}, \bibinfo
  {author} {\bibfnamefont {R.~E.}\ \bibnamefont {Robson}}, \bibinfo {author}
  {\bibfnamefont {M.~J.}\ \bibnamefont {Brunger}}, \ and\ \bibinfo {author}
  {\bibfnamefont {R.~D.}\ \bibnamefont {White}},\ }\href {\doibase
  10.1063/1.4885357} {\bibfield  {journal} {\bibinfo  {journal} {The Journal of
  chemical physics}\ }\textbf {\bibinfo {volume} {141}},\ \bibinfo {pages}
  {014308} (\bibinfo {year} {2014})}\BibitemShut {NoStop}%
\bibitem [{\citenamefont {White}\ \emph {et~al.}(2014)\citenamefont {White},
  \citenamefont {Brunger}, \citenamefont {Garland}, \citenamefont {Robson},
  \citenamefont {Ness}, \citenamefont {Garcia}, \citenamefont {de~Urquijo},
  \citenamefont {Dujko},\ and\ \citenamefont {Petrovi{\'{c}}}}]{White2014}%
  \BibitemOpen
  \bibfield  {author} {\bibinfo {author} {\bibfnamefont {R.~D.}\ \bibnamefont
  {White}}, \bibinfo {author} {\bibfnamefont {M.~J.}\ \bibnamefont {Brunger}},
  \bibinfo {author} {\bibfnamefont {N.~a.}\ \bibnamefont {Garland}}, \bibinfo
  {author} {\bibfnamefont {R.~E.}\ \bibnamefont {Robson}}, \bibinfo {author}
  {\bibfnamefont {K.~F.}\ \bibnamefont {Ness}}, \bibinfo {author}
  {\bibfnamefont {G.}~\bibnamefont {Garcia}}, \bibinfo {author} {\bibfnamefont
  {J.}~\bibnamefont {de~Urquijo}}, \bibinfo {author} {\bibfnamefont
  {S.}~\bibnamefont {Dujko}}, \ and\ \bibinfo {author} {\bibfnamefont {Z.~L.}\
  \bibnamefont {Petrovi{\'{c}}}},\ }\href {\doibase 10.1140/epjd/e2014-50085-7}
  {\bibfield  {journal} {\bibinfo  {journal} {The European Physical Journal D}\
  }\textbf {\bibinfo {volume} {68}},\ \bibinfo {pages} {125} (\bibinfo {year}
  {2014})}\BibitemShut {NoStop}%
\bibitem [{\citenamefont {{Koizumi}}\ \emph {et~al.}(1986)\citenamefont
  {{Koizumi}}, \citenamefont {{Shirakawa}},\ and\ \citenamefont
  {{Ogawa}}}]{koizumi86}%
  \BibitemOpen
  \bibfield  {author} {\bibinfo {author} {\bibfnamefont {T.}~\bibnamefont
  {{Koizumi}}}, \bibinfo {author} {\bibfnamefont {E.}~\bibnamefont
  {{Shirakawa}}}, \ and\ \bibinfo {author} {\bibfnamefont {I.}~\bibnamefont
  {{Ogawa}}},\ }\href@noop {} {\bibfield  {journal} {\bibinfo  {journal}
  {J.~Phys.~B: At. Mol. Phys.}\ }\textbf {\bibinfo {volume} {19}},\ \bibinfo
  {pages} {2331} (\bibinfo {year} {1986})}\BibitemShut {NoStop}%
\bibitem [{\citenamefont {{Pack}}\ \emph {et~al.}(1992)\citenamefont {{Pack}},
  \citenamefont {{Voshall}}, \citenamefont {{Phelps}},\ and\ \citenamefont
  {{Kline}}}]{pack92a}%
  \BibitemOpen
  \bibfield  {author} {\bibinfo {author} {\bibfnamefont {J.~L.}\ \bibnamefont
  {{Pack}}}, \bibinfo {author} {\bibfnamefont {R.~E.}\ \bibnamefont
  {{Voshall}}}, \bibinfo {author} {\bibfnamefont {A.~V.}\ \bibnamefont
  {{Phelps}}}, \ and\ \bibinfo {author} {\bibfnamefont {L.~E.}\ \bibnamefont
  {{Kline}}},\ }\href@noop {} {\bibfield  {journal} {\bibinfo  {journal} {J.
  Appl. Phys.}\ }\textbf {\bibinfo {volume} {71}},\ \bibinfo {pages} {5363}
  (\bibinfo {year} {1992})}\BibitemShut {NoStop}%
\bibitem [{\citenamefont {Boyle}\ \emph {et~al.}(2012)\citenamefont {Boyle},
  \citenamefont {White}, \citenamefont {Robson}, \citenamefont {Dujko},\ and\
  \citenamefont {{Lj Petrovi{\'{c}}}}}]{Boyle2012}%
  \BibitemOpen
  \bibfield  {author} {\bibinfo {author} {\bibfnamefont {G.~J.}\ \bibnamefont
  {Boyle}}, \bibinfo {author} {\bibfnamefont {R.~D.}\ \bibnamefont {White}},
  \bibinfo {author} {\bibfnamefont {R.~E.}\ \bibnamefont {Robson}}, \bibinfo
  {author} {\bibfnamefont {S.}~\bibnamefont {Dujko}}, \ and\ \bibinfo {author}
  {\bibfnamefont {Z.}~\bibnamefont {{Lj Petrovi{\'{c}}}}},\ }\href {\doibase
  10.1088/1367-2630/14/4/045011} {\bibfield  {journal} {\bibinfo  {journal}
  {New Journal of Physics}\ }\textbf {\bibinfo {volume} {14}},\ \bibinfo
  {pages} {045011} (\bibinfo {year} {2012})}\BibitemShut {NoStop}%
\bibitem [{\citenamefont {Yarnell}\ \emph {et~al.}(1973)\citenamefont
  {Yarnell}, \citenamefont {Katz}, \citenamefont {Wenzel},\ and\ \citenamefont
  {Koenig}}]{Yarnell73}%
  \BibitemOpen
  \bibfield  {author} {\bibinfo {author} {\bibfnamefont {J.}~\bibnamefont
  {Yarnell}}, \bibinfo {author} {\bibfnamefont {M.}~\bibnamefont {Katz}},
  \bibinfo {author} {\bibfnamefont {R.}~\bibnamefont {Wenzel}}, \ and\ \bibinfo
  {author} {\bibfnamefont {S.}~\bibnamefont {Koenig}},\ }\href@noop {}
  {\bibfield  {journal} {\bibinfo  {journal} {Phys. Rev. A}\ }\textbf {\bibinfo
  {volume} {7}},\ \bibinfo {pages} {2130} (\bibinfo {year} {1973})}\BibitemShut
  {NoStop}%
\bibitem [{\citenamefont {Shibamura}\ \emph {et~al.}(1984)\citenamefont
  {Shibamura}, \citenamefont {Masuda},\ and\ \citenamefont
  {Doke}}]{Shibetal1984}%
  \BibitemOpen
  \bibfield  {author} {\bibinfo {author} {\bibfnamefont {E.}~\bibnamefont
  {Shibamura}}, \bibinfo {author} {\bibfnamefont {K.}~\bibnamefont {Masuda}}, \
  and\ \bibinfo {author} {\bibfnamefont {T.}~\bibnamefont {Doke}},\ }in\
  \href@noop {} {\emph {\bibinfo {booktitle} {8th Workshop on Electron
  Swarms}}}\ (\bibinfo {year} {1984})\BibitemShut {NoStop}%
\bibitem [{\citenamefont {{Huang}}\ and\ \citenamefont
  {{Freeman}}(1978)}]{HuangFreem78}%
  \BibitemOpen
  \bibfield  {author} {\bibinfo {author} {\bibfnamefont {S.~S.~S.}\
  \bibnamefont {{Huang}}}\ and\ \bibinfo {author} {\bibfnamefont {G.~R.}\
  \bibnamefont {{Freeman}}},\ }\href@noop {} {\bibfield  {journal} {\bibinfo
  {journal} {J.~Chem.~Phys.}\ }\textbf {\bibinfo {volume} {47}},\ \bibinfo
  {pages} {1355} (\bibinfo {year} {1978})}\BibitemShut {NoStop}%
\end{thebibliography}%

\end{document}